\documentclass[12pt]{article}
\usepackage{amsfonts,amsmath,amssymb,mathrsfs}

\title{On Spontaneous Wave Function Collapse and
       Quantum Field Theory}
\author{ 
Roderich Tumulka\footnote{Mathematisches Institut,
    Eberhard-Karls-Unversit\"at, Auf der Morgenstelle 10, 72076
    T\"ubingen, Germany.  E-mail:
    tumulka@everest.mathematik.uni-tuebingen.de}
}
\date{December 14, 2005}

\newcommand{\RRR}{\mathbb{R}}

\newcommand{\PPP}{\mathbb{P}}
\newcommand{\Hilbert}{\mathscr{H}}
\renewcommand{\sp}[2]{\langle #1 | #2 \rangle}
\newcommand{\vx}{\boldsymbol{x}}
\newcommand{\vy}{\boldsymbol{y}}
\newcommand{\vr}{\boldsymbol{r}}

\newcommand{\vX}{{\boldsymbol{X}}}

\newcommand{\D}{\mathrm{d}} 
\newcommand{\E}{\mathrm{e}} 
\newcommand{\I}{\mathrm{i}} 

\newcommand{\K}{K}
\newcommand{\nconst}{\mathcal{N}}
\newcommand{\mconst}{\mathcal{M}}

\newcommand{\z}[1]{{#1}}

\begin{document}\maketitle\sloppy
\begin{abstract}
One way of obtaining a version of quantum mechanics without observers, and thus of solving the paradoxes of quantum mechanics, is to modify the Schr\"odinger evolution by implementing spontaneous collapses of the wave function. An explicit model of this kind was proposed in 1986 by Ghirardi, Rimini, and Weber (GRW), involving a nonlinear, stochastic evolution of the wave function. We point out how, by focussing on the essential mathematical structure of the GRW model and a clear ontology, it can be generalized to (regularized) quantum field theories in a simple and natural way.

\medskip

\end{abstract}

\section{Introduction}

John S.~Bell concluded from the quantum measurement problem that ``either the wave function, as given by the Schr\"odinger equation, is not everything or it is not right'' \cite{Belljumps}. Let us assume, for the purpose of this paper, the second option of the alternative: that the Schr\"odinger equation should be modified in such a way that superpositions of macroscopically different states, as exemplified by Schr\"odinger's cat, either cannot arise or cannot persist for more than a fraction of a second. Theories of this kind have come to be known under the names of ``dynamical reduction'', ``spontaneous localization'', or ``spontaneous wave function collapse,'' and have been advocated and studied by various authors \cite{BB66,Pe76,Pe79,Gi84,grw, Belljumps, diosi, pearle90, Pen00, adler, leggett, dowker1}; see \cite{BG03} for an overview. The merit of such theories, which I shall call ``collapse theories'' in the following, is that they provide, instead of statements about what observers would see if they were to make certain experiments, a possible story about what events objectively occur: they are, in other words, quantum theories without observers. In collapse theories, observations are merely special cases of the objective events, for which the theory, if it is to be empirically adequate, predicts the same distribution of outcomes as the standard quantum formalism. An example of a no-collapse quantum theory without observers is Bohmian mechanics \cite{DGZ04,Bellbook}.

An explicit collapse theory has been proposed by Ghirardi, Rimini, and Weber (GRW) in 1986 \cite{grw} for nonrelativistic quantum mechanics of $N$ distinguishable particles. While the testable predictions of the GRW model differ in principle from the quantum formulas, as yet no experiment could decide between the two, \z{provided one chooses the collapse rates proportional to the masses,} since the differences are too tiny for standard experiments and the experiments leading to noticeable differences are hard to carry out \cite{BG03}. \z{In the GRW model, the wave function obeys the unitary Schr\"odinger evolution until, at an unforeseeable, random time, it changes discontinuously in an unforeseeable, random way---it collapses. In order to obtain similar models for quantum field theories (QFTs), one path of research, which has been followed under the name ``continuous spontaneous localization'' (CSL) \cite{Pe89,GPR90}, is based on the idea of incessant mild collapses, so that the quantum state vector follows a diffusion process in Hilbert space. For a CSL model for identical particles, see \cite{Pe89}.}

I will show how \z{the GRW model can be extended to QFT in a more direct way; the resulting collapse QFT} is as hard to distinguish experimentally from standard QFT as the GRW model from standard quantum mechanics, while solving the quantum measurement problem in the same way as the GRW model. My proposal retains (and indeed is based on) the discreteness of the GRW model; it gets along with indistinguishable particles, both fermions and bosons, and with particle creation and annihilation. However, I will not try here to make the theory Lorentz-invariant; I hope to be able in a future work to combine the construction I present here with the Lorentz-invariant version of the GRW model for $N$ particles that I have described recently \cite{Tum05}. \z{When applied to a system of $N$ identical particles, my proposal yields a collapse process proposed already in 1995 by Dove and Squires \cite{dovethesis,DS95}, which, however, seems to have received little or no attention so far.}

\section{Mathematical Framework of the GRW Model}

I now describe the GRW model in an unusual way that emphasizes the abstract mathematical structure it is based on and uses an ontology proposed by Bell \cite{Belljumps,Bellexact}. According to this \emph{flash ontology}, matter consists of millions of so-called \emph{flashes}, physical events that are mathematically represented by space-time points. The flashes can be thought of as replacing the continuous particle trajectories in space-time postulated by classical mechanics. The flashes are random, thus forming what probabilists would call a point process in space-time, with a distribution determined by the (initial) wave function. The reader may wonder why, in a collapse theory, there is any need at all to introduce space-time objects such as flashes, a question that I will take up in Section~\ref{sec:collapses}. For now I ask for patience and suggest to regard it as the sole role of the wave function to determine the distribution of the flashes.

We begin with an (arbitrary) Hilbert space $\Hilbert$ with scalar product $\sp{\phi}{\psi}$, which in the GRW model is $L^2(\RRR^{3N})$, and an (arbitrary) self-adjoint operator $H$ on $\Hilbert$, the Hamiltonian, which in the GRW model is the usual Schr\"odinger operator $-\tfrac{\hbar^2}{2} \Delta +V$.  The third mathematical object that we need, besides $\Hilbert$ and $H$, is a 
\begin{equation}
  \text{positive-operator-valued function }\Lambda(\vx) 
\end{equation}
on physical space $\RRR^3$ acting on $\Hilbert$. (For mathematicians I should add the postulate that there is a dense domain in $\Hilbert$ on which all of the $\Lambda(\vx)$ are defined.) $\Lambda$ forms the link between Hilbert space and physical 3-space and can be thought of as representing the (smeared-out) position observable, and in particular as representing a preferred basis. In the GRW model, $\Lambda(\vx)$ is a multiplication operator
\begin{equation}\label{LambdaGRW}
  \Lambda(\vx) \, \psi(\vr_1, \ldots, \vr_N) = \nconst \, 
  \E^{-(\vx - \vr_i)^2/a^2} \, \psi(\vr_1, \ldots, \vr_N) \,,
\end{equation}
multiplying by a Gaussian with width $a/\sqrt{2}$ (in Bell's \cite{Belljumps} notation) \z{and center $\vx$, where $\nconst$ and $a$ are new constants of nature with suggested values of} 
\begin{equation}
  \nconst \approx 10^5\, \mathrm{sec}^{-1}\mathrm{m}^{-3}, \quad 
  a \approx 10^{-7} \, \mathrm{m} \,.
\end{equation}
\z{We are considering for the moment only flashes correspoding to the $i$-th particle coordinate $\vr_i$, $i \in \{1,\ldots,N\}$, and will later consider flashes due to all particles.}

The role of $\Lambda$ is to define the rate of a flash to occur, i.e., the probability per unit time. That is why I will call $\Lambda(\vx)$ the \emph{flash rate operators}. The rate at time $t=0$ of a flash in the set $B \subseteq \RRR^3$ is
\begin{equation}\label{rateB}
  \sp{\psi}{\Lambda(B)|\psi} \,,
\end{equation}
where $\Lambda(B)$ is short for
\begin{equation}
  \Lambda(B) = \int_B \D^3 \vx \, \Lambda(\vx) \,.
\end{equation}
This shows \z{why} $\Lambda(B)$ must have dimension $1/$time. In the GRW model, the total flash rate is independent of the quantum state $\psi$ since $\Lambda(\RRR^3) = \pi^{3/2}\nconst a^3 I$ is a multiple of the identity operator $I$ on $\Hilbert$. The constant in front of $I$, or, equivalently, the total flash rate, is called $1/\tau$ in Bell's \cite{Belljumps} notation, with
\begin{equation}
  \tau \approx 10^{15} \, \mathrm{sec} \,.
\end{equation}
\z{It has been pointed out by Pearle and Squires \cite{PS94} that a fixed $\tau$ would contradict the observed stability of nucleons; they suggest instead that the flash rate constant $\nconst a^3$ be proportional to the mass of a particle, rather than a universal constant.} In the QFT model we will devise, \z{however,} $\Lambda(\RRR^3)$ will not be a multiple of the identity, and this is the only aspect in which we essentially generalize the mathematical structure of the GRW model.

We now define the probability distribution of the first flash $(\vX_1,T_1)$ with random location $\vX_1$ and random time $T_1$, a probability distribution on the space-time region with $t>0$. The distribution is quadratic in $\psi$. The wish that the rate be given by $\Lambda$ and that the evolution before the flash be given essentially by $H$ leads us to the following form for the distribution:
\begin{equation}\label{1flashdist}
  \PPP\bigl( \vX_1 \in \D^3 \vx, T_1 \in \D t \bigr) =
  \sp{\psi}{W_t^* \Lambda(\vx) W_t|\psi} \, \D^3\vx \, \D t \,,
\end{equation}
where the star denotes the adjoint operator, and 
\begin{equation}\label{Wdef}
  W_t = \E^{-\frac{1}{2} \Lambda(\RRR^3)t - \frac{\I}{\hbar} Ht} 
  \text{ for }t\geq 0\,, \quad W_t = 0 \text{ for } t<0 \,.
\end{equation}
Without the $\Lambda(\RRR^3)$ term, this would be the ordinary unitary evolution; we need the additional term to keep track of the probability that time $t$ is reached without a flash. Indeed, the definition \eqref{Wdef} implies that \eqref{1flashdist} is a probability distribution:
\begin{subequations}\label{calculation}
\begin{align}
  &\int\D^3 \vx \int\limits_0^\infty \D t \, \sp{\psi}{W_t^* \Lambda(\vx) W_t|\psi} =
  \int\limits_0^\infty \D t \, \sp{\psi}{W_t^* \Lambda(\RRR^3) W_t|\psi} = \\
  &= -\int\limits_0^\infty \D t \, \sp{\psi}{W_t^* \Bigl(- \tfrac{1}{2} \Lambda(\RRR^3) + 
  \tfrac{\I}{\hbar} H- \tfrac{1}{2} \Lambda(\RRR^3) - 
  \tfrac{\I}{\hbar} H \Bigr) W_t|\psi} =\\
  &= -\int\limits_0^\infty \D t \, \sp{\psi}{ (\dot{W}_t^* W_t + 
  W_t^* \dot{W}_t ) |\psi} =\\
  &= -\int\limits_0^\infty \D t \, \frac{\D}{\D t} \sp{\psi}{W_t^* W_t |\psi} =
  \sp{\psi}{W_0^* W_0|\psi} = \sp{\psi}{\psi} = 1 \,,
\end{align}
\end{subequations}
provided $W_t \to 0$ as $t \to \infty$, which is the case if the spectrum of $\Lambda(\RRR^3)$ is bounded away from zero  (as in the GRW case).\footnote{Otherwise, the theory is not ill-defined, but rather the flash rate is so low for some $\psi$ that there is positive probability that no flash occurs.} The same calculation for a time integral from 0 to $t$ shows that the probability of a flash before $t$ equals $1-\|W_t \psi\|^2$; in particular we can see why $W_t$ should not be unitary. In the GRW case, since $\Lambda(\RRR^3)$ is a multiple of the identity and thus commutes with $H$, we find that the exponential \eqref{Wdef} splits into a product of two exponentials,
\begin{equation}
  W_t = \E^{-t/2\tau} \E^{-\I Ht/\hbar} \text{ for }t \geq 0\,. 
\end{equation}

The joint distribution of the first $n$ flashes is defined to be
\begin{multline}\label{nflashdist}
  \PPP\bigl( \vX_1\in \D^3 \vx_1, T_1 \in \D t_1, \ldots,
  \vX_n \in \D^3 \vx_n, T_n \in \D t_n \bigr) =\\ 
  \bigl\| \K_n(0,\vx_1, t_1, \ldots, \vx_n, t_n) \, \psi \bigr\|^2 
  \, \D^3 \vx_1 \, \D t_1 \cdots \D^3 \vx_n \, \D t_n \,,
\end{multline}
where $\K_n$ is an operator-valued function on (space-time)$^n$ defined by
\begin{multline}\label{Kndef}
  \K_n(t_0,\vx_1, t_1, \ldots, \vx_n, t_n) =\\
  \Lambda(\vx_n)^{1/2} \,W_{t_n-t_{n-1}} \Lambda(\vx_{n-1})^{1/2} \,W_{t_{n-1}-t_{n-2}}
  \cdots \Lambda(\vx_1)^{1/2} \, W_{t_1-t_0} \,.
\end{multline}
The square-roots exist since the $\Lambda(\vx)$ are positive operators. Observing that, by a reasoning analogous to \eqref{calculation}, 
\begin{equation}
  \int \D^3 \vx_n \int\limits_{t_{n-1}}^\infty \D t_n \, \K^*_n \,\K_n = 
  \K^*_{n-1} \, \K_{n-1} \,,
\end{equation}
we see two things: firstly that the right hand side of \eqref{nflashdist} is a probability distribution on $(\text{space-time})^n$, and secondly that these distributions, for different values of $n$, are marginals of each other, thus forming a consistent family and arising from a joint distribution of infinitely many random variables $\vX_1, T_1, \vX_2, T_2, \ldots$. 

The original GRW model contains one further complication: that collapses can act on different coordinates, as encoded in the particle index $i$ in \eqref{LambdaGRW}. Reflecting the fact that the model should account for the quantum mechanics of $N$ \emph{distinguishable} particles, we simply postulate that there are $N$ different \emph{types of flashes}, or, equivalently, that each flash is labeled by an index $i \in \{1,\ldots,N\}$. Correspondingly, we need to be given $N$ positive-operator-valued functions $\Lambda_i(\vx)$, while $\Hilbert$ and $H$ are the same for all types of flashes. Thus, with every flash $(\vX_k,T_k)$ is associated a random label $I_k \in \{1, \ldots, N\}$, and the joint distribution is defined to be
\begin{multline}\label{Nnflashdist}
  \PPP\bigl( \vX_1\in \D^3 \vx_1, T_1 \in \D t_1, I_1 = i_1, \ldots,
  \vX_n \in \D^3 \vx_n, T_n \in \D t_n, I_n = i_n \bigr) =\\ 
  \bigl\| \K_n(0,\vx_1, t_1,i_1, \ldots, \vx_n, t_n,i_n) \, \psi \bigr\|^2 
  \, \D^3 \vx_1 \, \D t_1 \cdots \D^3 \vx_n \, \D t_n \,,
\end{multline}
where $\K_n$ is now an operator-valued function on $\bigl[(\text{space-time}) \times \{1,\ldots,N\}\bigr]^n$ defined by
\begin{multline}\label{Knidef}
  \K_n(t_0,\vx_1, t_1,i_1, \ldots, \vx_n, t_n, i_n) =\\
  \Lambda_{i_n}(\vx_n)^{1/2} \,W_{t_n-t_{n-1}} \Lambda_{i_{n-1}}(\vx_{n-1})^{1/2} 
  \,W_{t_{n-1}-t_{n-2}}
  \cdots \Lambda_{i_1}(\vx_1)^{1/2} \, W_{t_1-t_0} \,,
\end{multline}
with
\begin{equation}\label{WNdef}
  W_t = \exp\Bigl(-\tfrac{1}{2} \sum_{i=1}^N \Lambda_i (\RRR^3)t 
  - \tfrac{\I}{\hbar} Ht\Bigr) 
  \text{ for }t\geq 0\,, \quad W_t = 0 \text{ for } t<0 \,.
\end{equation}
In the same way as before, one checks that
\begin{equation}
  \sum_{i_n=1}^N \int \D^3 \vx_n \int\limits_{t_{n-1}}^\infty \D t_n \, \K^*_n \,\K_n = 
  \K^*_{n-1} \, \K_{n-1}
\end{equation}
and
\begin{equation}
  \sum_{i_1=1}^N \int \D^3 \vx_1 \int\limits_{0}^\infty \D t_1 \, \K^*_1 \,\K_1 = I \,,
\end{equation}
implying that \eqref{Nnflashdist} is a consistent family of probability distributions. This completes our definition of a theory from $\Hilbert$, $H$, and $\Lambda_1(\vx), \ldots, \Lambda_N(\vx)$, including the GRW model.

\section{QFT}

Since $\Hilbert$ can be taken to be the Hilbert space of a QFT and $H$ its (regularized) Hamiltonian, we get a collapse version of that QFT as soon as we have the flash rate operators $\Lambda(\vx)$. The model is then defined by \eqref{Wdef}, \eqref{nflashdist}, and \eqref{Kndef}. If the QFT contains several species of particles, we may wish to introduce one type of flash for every species, and use eq.s \eqref{Nnflashdist}, \eqref{Knidef}, and \eqref{WNdef} instead.

So what would be a natural choice of $\Lambda(\vx)$? Since the operator $\Lambda(B)$ determines the flash rate in $B$, it should represent the amount of matter in $B$, smeared out by a Gaussian with width $a/\sqrt{2}$. \z{Two natural choices are the particle number density operator $N(\vy)$ and the mass density operator $M(\vy)$, that is,}
\begin{equation}\label{LambdaN}
  \Lambda(\vx) = \int \D^3 \vy \, \nconst \, \E^{-(\vx-\vy)^2/a^2} \, N(\vy)\,,
\end{equation}
\z{respectively}
\begin{equation}\label{LambdaM}
  \Lambda(\vx) = \int \D^3 \vy \, \mconst \, \E^{-(\vx-\vy)^2/a^2} \, M(\vy)\,,
\end{equation}
\z{with $\mconst$ a suitable constant. The operators \eqref{LambdaN} have been considered already in \cite{GPR90} for CSL in connection with identical particles, albeit apparently in a role more analogous to $\Lambda(\vx)^{1/2}$ than to $\Lambda(\vx)$. The choice of \eqref{LambdaM} instead of \eqref{LambdaN} corresponds to the proposal of Pearle and Squires \cite{PS94} to choose the collapse rate proportional to the mass.}

\z{In general, the number operator $N(\vy)$ can be expressed in terms of suitable annihilation operators $a(\vy)$ and creation operators $a^*(\vy)$ in the position representation, acting on a suitable Fock space $\Hilbert$, by} 
\begin{equation}
  N(\vy) = a^*(\vy) \, a(\vy), 
\end{equation}
\z{where the product involves, when appropriate, summation over spin indices. In a nonrelativistic QFT, $a(\vy)$ is simply the field operator at the location $\vy$. (Of course, since in nonrelativistic quantum theories usually the Hamiltonian does not contain terms creating and annihilating particles, we would have to add such terms artificially to the Hamiltonian to obtain a model with non-conserved particle number.)}

For several species \z{of particles} corresponding to several \z{quantum} fields, we \z{thus} obtain several rate density operators $\Lambda_i(\vx)$.

\section{Second Quantization}

I would like to describe another way of constructing flash rate operators $\Lambda(\vx)$ for QFT. It will turn out equivalent to \eqref{LambdaN}. It is based on the second quantization algorithm for forming a (bosonic or fermionic) Fock space $\Hilbert$ from a one-particle Hilbert space $\Hilbert_{(1)}$, and consists of an algorithm for forming flash rate operators $\Lambda(\vx)$ acting on Fock space $\Hilbert$ from flash rate operators $\Lambda_{(1)}(\vx)$ acting on $\Hilbert_{(1)}$. This algorithm in turn is based on two procedures, one concerning direct sums of Hilbert spaces and the other tensor products.

On the direct sum $\Hilbert_{1} \oplus \Hilbert_{2}$ of two Hilbert spaces, each equipped with a positive-operator-valued function $\tilde\Lambda_{i}(\vx)$, $i=1,2$, the natural way of obtaining a positive-operator-valued function is
\begin{equation}\label{Lambdasum}
  \Lambda(\vx) = \tilde\Lambda_{1}(\vx) \oplus \tilde\Lambda_{2}(\vx) \,.
\end{equation}

On the tensor product space $\Hilbert_{1} \otimes \Hilbert_{2}$, it is natural to consider the two functions 
\begin{equation}\label{2Lambdas}
  \Lambda_1 (\vx) = \tilde\Lambda_{1}(\vx) \otimes I_{2}\text{ and }
  \Lambda_2(\vx) = I_{1} \otimes \tilde\Lambda_{2}(\vx)\,, 
\end{equation}
defining a theory with two types of flashes. A relevant property of this choice is that if the two physical systems corresponding to $\Hilbert_{1}$ and $\Hilbert_{2}$ do not interact, $H = H_{1} \otimes I_{2} + I_{1} \otimes H_{2}$, and if the initial state vector factorizes, $\psi = \psi_{1} \otimes \psi_{2}$, then type-1 and type-2 flashes are independent of each other. Indeed, the type-1 flashes are also independent of $H_{2}$, $\Lambda_{2}$, and $\psi_{2}$, and vice versa. Each of the two systems behaves as if it was alone in the world, obeying its own version of the law \eqref{nflashdist}, and that is a reasonable behavior. 

Suppose we do not want two types of flashes, but one. Observe that, \z{by \eqref{rateB}, the rate (at time $t=0$) for a flash \emph{of any type} to occur in a set $B\subseteq \RRR^3$ is $\langle \psi |\Lambda_1(B)|\psi \rangle + \langle \psi |\Lambda_2(B)|\psi \rangle$, the same as the rate of flashes of only one type with rate operators}
\begin{equation}\label{Lambdaprod}
  \Lambda(\vx) = \tilde\Lambda_{1}(\vx) \otimes I_{2} + 
  I_{1} \otimes \tilde\Lambda_{2}(\vx) \,,
\end{equation}
which is thus a natural choice of a positive-operator-valued function on the tensor product.

On the $N$-th tensor power $\Hilbert_{(1)}^{\otimes N}$ \z{of the one-particle Hilbert space $\Hilbert_{(1)}$}, the \z{flash rate operator corresponding to \eqref{Lambdaprod},}
\begin{equation}
  \Lambda(\vx) = \sum_{i=1}^N I^{\otimes (i-1)} \otimes \Lambda_{(1)}(\vx) 
  \otimes I^{\otimes (N-i)}\,,
\end{equation}
can be written using the permutation operators $U_\sigma$ on $\Hilbert_{(1)}^{\otimes N}$ for $\sigma$ in the permutation group $S_N$,
\begin{equation}\label{symmLambda}
  \Lambda(\vx) = \frac{1}{(N-1)!} \sum_{\sigma \in S_N} U_\sigma^* 
  \bigl( \Lambda_{(1)}(\vx) 
  \otimes I^{\otimes (N-1)} \bigr) U_\sigma \,,
\end{equation}
and therefore assumes values in the symmetric operators, mapping in particular  symmetric (bosonic) vectors to symmetric ones and anti-symmetric (fermionic) vectors to anti-symmetric ones, thus defining two positive-operator-valued functions $\Lambda_{(N)}^{\pm}(\vx)$ acting on the bosonic ($+$) respectively fermionic ($-$) $N$-particle Hilbert space: \z{$\Lambda_{(N)}^{\pm}(\vx)$ is the restriction of the $\Lambda(\vx)$ given by \eqref{symmLambda} to the bosonic respectively fermionic subspace of $\Hilbert_{(1)}^{\otimes N}$. The flash theory for $N$ bosons or fermions with these flash rate operators is closely related, in a way that will become clear in the next section, to the collapse process proposed by Dove and Squires \cite{dovethesis,DS95}.}

\z{Adding the} $\Lambda^\pm_{(N)}(\vx)$ in the sense of \eqref{Lambdasum} from $N=0$ to $\infty$ yields two functions $\Lambda^{\pm}(\vx)$ acting on the bosonic respectively fermionic Fock space. This completes our construction for the ``second quantization" of $\Lambda_{(1)}(\vx)$. If we take $\Lambda_{(1)}(\vx)$ to be the multiplication operator 
\eqref{LambdaGRW} with $N=1$ (and $i=1$) and $a(\vx)$ the canonical annihilation operator on Fock space, then $\Lambda(\vx)$ coincides with \eqref{LambdaN}.

\section{Collapses}\label{sec:collapses}

After talking so much about flashes, I should point out what they have to do with collapses of the wave function. Suppose that $n$ flashes have occurred between time $0$ and time $t$, with the $k$-th flash at time $t_k$ and location $\vx_k$. Then the distribution of the next $m$ flashes after time $t$, conditional on the history of flashes between $0$ and $t$, is, as a consequence of  \eqref{nflashdist}, given by
\begin{multline}\label{condflashdist}
  \PPP \Bigl( \vX_{n+1} \in \D^3 \vx_{n+1}, T_{n+1} \in \D t_{n+1}, \ldots,
  \vX_{n+m} \in \D^3 \vx_{n+m}, T_{n+m} \in \D t_{n+m} \Big|\\
  \vX_1 = \vx_1, T_1 = t_1, \ldots, \vX_n = \vx_n, T_n = t_n\leq t, T_{n+1}\geq t \Bigr) =\\
  \bigl \| \K_m(t, \vx_{n+1}, t_{n+1}, \ldots, \vx_{n+m},t_{n+m}) \psi_t \|^2 \,
  \D^3 \vx_{n+1} \, \D t_{n+1} \cdots \D^3 \vx_{n+m} \, \D t_{n+m} \,,
\end{multline}
where $\psi_t$ is the \emph{conditional wave function}
\begin{equation}\label{condpsidef}
  \psi_t = \frac{ W_{t-t_n} \, \K_n (0,\vx_1,t_1, \ldots, \vx_n,t_n) \, \psi }
  {\| W_{t-t_n} \, \K_n (0,\vx_1,t_1, \ldots, \vx_n,t_n) \, \psi \|} \,.
\end{equation}
\z{As a corollary, the flash rate at time $t$ in a set $B \subseteq \RRR^3$ is}
\begin{equation}\label{rateBt}
  \langle \psi_t | \Lambda (B)| \psi_t \rangle\,.
\end{equation}
It is $\psi_t$ that collapses whenever a flash occurs, say at $(\vX,T)$, according to
\begin{equation}\label{collapse}
  \psi_{T+} = \frac{\Lambda(\vX)^{1/2} \, \psi_{T-}} 
  {\|\Lambda(\vX)^{1/2} \, \psi_{T-} \|} \,,
\end{equation}
and evolves deterministically between the flashes according to the operators $W_t$ (up to normalization). \z{More explicitly, for the example case of $N$ identical particles, the (anti)symmetric wave function $\psi(\vr_1, \ldots, \vr_N)$ collapses to (a normalization factor times)}
\begin{equation}
  \Lambda_{(N)}^{\pm}(\vX)^{1/2} \, \psi(\vr_1, \ldots, \vr_N) = \biggl(
  \nconst \sum_{i=1}^N  
  \E^{-(\vX - \vr_i)^2/a^2} \biggr)^{1/2} \, \psi(\vr_1, \ldots, \vr_N)
\end{equation}
\z{for bosons $(+)$ respectively fermions $(-)$. It is this formula for the collapsed wave function that Dove and Squires proposed \cite{DS95}.}

\z{Since the conditional wave function $\psi_t$ is random, one can form the density matrix}
\begin{equation}
  \rho_t = \int\limits_\Hilbert \PPP(\psi_t \in \D \phi) \, |\phi\rangle \langle \phi|  
\end{equation}
\z{of its distribution, in other words the density matrix of an ensemble of systems, each of which started with the same initial wave function $\psi$ but experienced flashes independently of the other systems. For the sake of completeness we note that it can be computed to be}
\begin{equation}
  \rho_t = \sum_{n=0}^\infty \int \D^3 \vx_1 \cdots \D^3\vx_n \int\limits_0^t \D t_1 
  \cdots \int\limits_{t_{n-1}}^t \D t_n \, W_{t-t_n} \, K_n |\psi \rangle \langle \psi | K_n^*
  \, W^*_{t-t_n}
\end{equation}
\z{with $K_n = K_n(0,\vx_1,t_1, \ldots, \vx_n,t_n)$, and obeys the master equation}
\begin{equation}
  \frac{\D \rho_t}{\D t} = -\tfrac{\I}{\hbar} [H,\rho_t] - \tfrac{1}{2} 
  \{\Lambda(\RRR^3), \rho_t \} + \int \D^3 \vx \, \Lambda (\vx)^{1/2} \,
  \rho_t \, \Lambda(\vx)^{1/2}
\end{equation}
\z{with $[\,,]$ the commutator and $\{\,,\}$ the anti-commutator.}

It is tempting to regard the collapsed wave function $\psi_t$ as the ontology, but I insist that the flashes form the ontology. This is a subtle point. After all, since for example the wave function of Schr\"odinger's cat, $(|\text{alive}\rangle + |\text{dead}\rangle)/\sqrt{2}$, quickly collapses into essentially either $|\text{alive}\rangle$ or $|\text{dead}\rangle$, it may seem that the collapsed wave function represents reality. However, that this view is problematic becomes evident when we note that even after the collapse into $\psi_t \approx |\text{alive}\rangle$, the coefficient of $|\text{dead}\rangle$ in $\psi_t$ is tiny but not zero. How small would it have to be to make the cat alive? The more fundamental problem with this view is that while the wave function may \emph{govern} the behavior of matter, it \emph{is} not matter; instead, matter corresponds to variables \emph{in space and time} \cite{AGTZ}, called ``local beables'' by Bell \cite{Bellbook} and ``primitive ontology'' by D\"urr, Goldstein, and Zangh\`\i~\cite{DGZ04}.

In what I described in the previous sections, the flashes form the primitive ontology. But other choices are possible, and this fact underlines that the theory is not completely specified by the stochastic evolution law for $\psi_t$ alone. An example of a different primitive ontology, instead of flashes, is the \emph{matter density ontology}, a continuous distribution of matter in space with density
\begin{equation}\label{mdef}
  m(\vx,t) = \sp{\psi_t}{\Lambda(\vx)|\psi_t}
\end{equation}
in our notation. While the two theories (using the same wave function with either the flash ontology or the matter density ontology) cannot be distinguished empirically, they differ metaphysically and physically. For example, the equations I considered in \cite{Tum05} define a Lorentz-invariant theory with the flash ontology but not with the matter density ontology, and strong superselection rules \cite{ssr} can hold with the flash ontology but not with the matter density ontology. For further discussion of the concept of primitive ontology see \cite{AGTZ}.

\section{Predictions}

The empirically testable predictions of the collapse QFT model we described agree with the standard predictive rules of QFT to the same extent as the GRW model \z{(say, with mass-dependent collapse rate)} agrees with standard quantum mechanics. To see this, recall first from the paragraph containing eq.~\eqref{2Lambdas} that when a system (defined, e.g., by a region in 3-space) can be regarded as decoupled and disentangled from its environment then its flash process is independent of the environment. Next note that, \z{by \eqref{rateBt}}, the total flash rate is $\sp{\psi_t}{\Lambda(\RRR^3)|\psi_t}$, proportional to \z{either} the (value regarded in quantum theory as the) average number of particles (relative to $\psi_t$) \z{or the average net mass}. As a consequence, as with the GRW model, a system containing fewer than a thousand particles experiences no more flashes than once in 100,000 years. Up to the first flash, the deviation of $\psi_t$ from the Schr\"odinger evolution is small for $t \ll \tau/\Delta N$ if $\Delta N$ is a bound on the spread in particle number of $\E^{-\I Hs/\hbar} \,\psi$, $0\leq s \leq t$. A macroscopic piece of matter, in contrast, with over $10^{22}$ particles, experiences millions of flashes every second. A macroscopic superposition essentially breaks down with the first flash (with consequences for the distribution of the future flashes) to one of the contributions, and for the same reasons as in the GRW model the random choice of the surviving contribution occurs with almost exactly the quantum theoretical probabilities. And like in the GRW model, this entails that experiments that ``measure'' any quantum ``observable'' on a microscopic system have almost exactly the quantum theoretical distribution of outcomes. \z{Dove and Squires \cite{DS95} provide a discussion of the consistency of their collapse process for identical particles with those experimental data that yield restrictions on the possibility of spontaneous collapse.}

Some collapse theories imply the possibility of superluminal (i.e., faster than light) signalling; even if the theory is hard to distinguish empirically from standard quantum theories, those experiments sensitive enough to detect the deviation can allow signalling using EPR--Bell pairs. Such collapse theories are therefore unlikely to possess a Lorentz-invariant version. In contrast, the collapse QFT developed here and the GRW model exclude superluminal signalling; this follows essentially from their property that the distribution of the flashes is quadratic in $\psi$. Indeed, if two systems are entangled but decoupled, $H = H_{1} \otimes I_{2} + I_{1} \otimes H_{2}$, and the flash rate operators are additive according to \eqref{2Lambdas} or \eqref{Lambdaprod}, then, as a consequence of \eqref{nflashdist}, the marginal distribution of the flashes belonging to system 1 depends on $H$, $\Lambda$, and $\psi$ only through $H_{1}$, $\tilde\Lambda_{1}$, and the reduced density matrix $\mathrm{tr}_{2} |\psi \rangle \langle \psi|$ of system 1. \z{To see this, consider first the case \eqref{2Lambdas}, in which there are two types of flashes for the two systems; in this situation the claim follows from the fact that the operators $\K_n$ defined by \eqref{Knidef} decompose into $\K_{1,n_{1}} \otimes \K_{2,n_{2}}$. In the case \eqref{Lambdaprod} of a single type of flashes it is not obvious which flashes are to be attributed to which system, unless the two systems have disjoint supports ($S_1,S_2 \subseteq \RRR^3$ with $S_1 \cap S_2 = \emptyset$) and we count the flashes in $S_1$ for system 1. But then we can in fact regard the flashes as labeled, the label being a function of the location, corresponding to $\Lambda_i(\vx) = 1_{S_i}(\vx) \, \Lambda(\vx)$ with $1_{S_i}$ the characteristic function of $S_i$, which brings us back to the previous case.}

\section{Literature}

GianCarlo Ghirardi sometimes suggests in his writings that identical particles or QFT cannot be treated in the framework of the GRW model in a satisfactory way \cite[page~118]{Ghi98}, \cite[pages 312 and 382]{BG03}, but require a diffusion process in Hilbert space; I think that the model I have presented \z{(respectively, as far as $N$ identical particles are concerned, the model of Dove and Squires \cite{DS95})} is a counterexample. \z{Part of the reason why the model of Dove and Squires has not received enough attention may be that they have not made clear enough, in my view, its naturalness and simplicity, and that they have presented it on equal footing with another, much less natural, proposal.}

I know of \z{five} variants of the GRW model for identical particles that have been proposed, \z{apart from the one discussed in this paper}: One was introduced by Ghirardi, Nicrosini, Rimini, and Weber in 1988 \cite{GNRW88}, which, however, appears theoretically unsatisfactory since it prescribes that the flashes of a system of $N$ identical particles occur in clusters of $N$ simultaneous ones, leaving no hope for a Lorentz-invariant version. It is also presumably empirically inadequate since it predicts that a superposition of two wave packets for the same single particle at a distance greater than 10 kilometers collapses within $10^{-7} \, \mathrm{sec}$. The second proposal was made by Kent in 1989 \cite{kent}, in which, however, the distribution of the flashes is not quadratic in $\psi$, thus allowing superluminal signalling. \z{The same is true of the third proposal, which is contained as well in the paper of Dove and Squires \cite{DS95}.} The \z{fourth} proposal was made by Bell in 1987, but never published, \z{except in a brief description in Sec.~IV~C of \cite{GPR90}}. I have learnt about it from Alberto Rimini on a recent conference in honor of GianCarlo Ghirardi's 70th birthday at Trieste and Mali Losinj; there I have learnt as well that equations similar to \eqref{Wdef} and \eqref{collapse} had already been considered in a different context, namely for models of coupling classical and quantum systems \cite{BJ95}. \z{The fifth variant was mentioned by Ghirardi, Pearle, and Rimini in 1990 as a side remark in Sec.~IV~C of \cite{GPR90}: The process for the wave function is similar in spirit to the proposal of \cite{GNRW88}, involving a simultaneous collapse for all particles, but differs in that it cannot be associated with flashes, as the collapse involves a continuous distribution function $n(\vx)$ on, rather than a set of $N$ points in, 3-space.}

\bigskip

\noindent\textit{Acknowledgments.} \z{I thank Philip Pearle for his detailed comments on a previous version of this article.}

\end{document}